\documentstyle[aps]{revtex}

\begin{document}
\title{Fast fluctuating fields as the source of low-frequency conductance
fluctuations in many-electron systems and failure of quantum kinetics}
\author{Yu.E. Kuzovlev}
\address{DonPTI NAS, Donetsk, Ukraine}
\date{April 20, 2000}

\begin{abstract}
It is shown that in many-electron systems quantum transfer amplitudes 
and thus transfer probabilities may be strongly influenced by fast 
fluctuating fields, in particular, caused by simultaneous electron 
transfers. Corresponding mutual interplay of many electron jumps,  
arising at the fundamental level of quantum phases, results in 
long-correlated (1/f type) conductance fluctuations. However, thats 
could not be theoretically catched if neglect the real discreteness of
quantum energy spectra and use the continuous spectrum approximation 
when building kinetic theory. Basing on first principles, the estimates 
of low-frequency fluctuations of tunneling conductance are presented.
\end{abstract}

\section{Introduction}

The electrical 1/f-noise (flicker noise) was discovered in 1925. Up to now
low frequency 1/f-noise is known in different physical, chemical,
biological, cosmic, geophysical and social systems [1-5], but there is no
conventional explanation of this phenomenon . In electronics frequently just
the 1/f conductance fluctuations limit qualities of real devices. What is
important, 1/f-noise is much higher sensitive to mesoscopic structure of
conductors as well as to external influences than conductivity itself,
hence, it may bring a rich and delicate information on microscopic
mechanisms of conductivity.

Traditionally 1/f-noise is thought be reducable to some ''slow'' random
processes with a broad distribution of large life-times (relaxation times).
The standard idea is that conductivity determined by ''fast'' mechanisms is
modulated by ''slow'' varying parameters (number of charge carriers, local
disorder, occupation of electron traps, Coulomb potentials, lattice magnetic
moments, etc., for some hypotheses see [1-4,6-14,17-18]). But concrete
origin of long life-times remains misterious even in best experimentally
investigated situations such as the electron mobility 1/f fluctuations in
semiconductors [1,5,10] (besides, this theory could scarcely likely say why
1/f-noises in liquid and solid metals are comparable [5], or why
hypothetical activation energies required to fit observed 1/f spectra much
exceed actual energies determined from the response to charge injections
[7]).

The problem looks sharply in case of ''bad'' (narrow-band, variable-range
hopping, tunnel, magnetically disturbed) conductivity like in doped
semiconductors [4,7,11], oxides [6,9], amorphous silicon [11-14], materials
with colossal magnetoresistance [15-16,19,20,32], cermets [36], etc. In
these systems long-range Coulombian (or magnetic) interactions are important
stimulating the attempts to reduce 1/f-noise to slow random charge
redistributions [4,11-14,17-18], but it is hard to explain why real 1/f
spectra do not saturate below 100 Hz [7,17,18]. The interesting concept of
self-organized criticality [31] also connects low-frequency noise with large
spatial and temporal scales.

The alternative way to understanding 1/f-noise is developed since 1982
[5,21-25,28-30]. It attributes 1/f-noise not to ''slow'' processes but, in
opposite, to ''fast'' microscopic kinetic processes responsible for
resistivity. The logics is simple: if producing stochastic behaviour any
dynamical system constantly forgets its own history, hence, it is
indifferent to a number of kinetic events happened in the past, hence, it
has neither stimulus nor possibility to follow some absolutely certain
''probability of event per unit time''. Therefore, a dynamical (Hamiltonian)
system what produces relaxation (irreversibility) and noise makes this,
generally speaking, without keeping definite (well time-averagable)
''probabilities per unit time'' and consequently produces fluctuations of
the relaxation (dissipation) and noise rates. These fluctuations do not
destroy detailed balance and do not cause any compensating reaction, hence,
have no characteristic time scale (see [25] for more details and examples).
In this theory the long-living statistical correlations associated with 1/f
spectrum are the manifestation of pure freedom of random flow of kinetic
events, not of long memory [21-25]. In this respect, as N.S.Krylov argued
[26] in 1950, we all should get rid of the erroneous opinion that
statistical correlations always reflect some actual causality.

The kinetic theory misses this 1/f-noise just because assumes quite certain
''probabilities per unit time'' although this ansatz is masked by ansatzes
like ''molecular chaos'', ''random phases'', ''thermodynamical limit'',
''continuous spectra'', etc. But the correctly performed derivation of gas
kinetics from Hamiltonian dynamics [28,25] demonstrates violation of
''molecular chaos'' and 1/f fluctuations of diffusivities and mobilities of
gas particles. Thus 1/f-noise arises even in the system where wittingly
nothing like slow processes and giant life-times do exist.

While a gas is characterized by strong but short (well time-space separated)
kinetic events (collisions), many systems possess weak but long-lasting (bad
separated) interactions. For example, phononic systems (dielectric
crystals). Nevertheless, as was shown in [30], 1/f type fluctuations of
dissipation (inner friction) and of Raman light scattering observed in these
systems (quartz) also can be deduced just from the basic (phonon-phonon)
kinetics, if only take into account that any particular interaction
interplays parametrically with other simultaneous interactions. Thus, it is
wrong to suppose ''elementary'' kinetic events be statistically independent,
in contrary to what we see in kinetics.

The purpose of the present paper is to show that similar situation may
realize in quantum conducting many-electron systems. Concretely, that the
very evolution of quantum amplitude and probability concerning one
particular electron transfer may depend on fast fluctuating potentials and
fields (electric, magnetic, etc.) induced by other simultaneously occuring
transfers, i.e. many elementary kinetic events (quantum jumps) are
essentially interplaying. As far as we know, this effect still was not under
consideration. It is expected be especially strong if characteristic
transfer time (the time of evolution of quantum transfer probability right
up to certainty) noticably exceeds correlation time of fluctuating fields.
Such a condition seems natural if any conducting electron feels
displacements of many other electrons (or changes of many magnetic moments)
by means of long-range interactions.

Our first principal statement is that short-correlated random fields if
influencing the formation of quantum transfer amplitudes lead to strong
uncertainty of transfer probabilities, thus well defined ''probabilities per
unit time'' do not exist. As a consequence, long-correlated conductance
fluctuations arise whose principal scale properties coinside with thats of
1/f-noise. Hence, to explain 1/f-noise we have no need in extremely slow
charge redistributions, instead the very fast ones are sufficient. The
second statement is that these effects can not be catched if neglect the
actual quantum discreteness of energy spectra. 

For simplicity, here to comprehend the main idea we concentrate on the case
of tunnel conduction (some aspects of the discreteness in 
tunnel junctions were touched in [37] but only relating to low temperatures 
without accounting for Coulombian effects). 
More extended argumentation including rigorous analysis of
Hamiltonian models will be published separately.

\section{Characteristic times of tunnel conductivity}

Let us consider characteristic time scales relating to electron tunneling
between metallic sides. Under a small voltage $U$ applied to tunnel junction
the mean charge transported during time $\Delta t$ can be phenomenologically
expressed as
\[
\Delta Q=e\cdot \frac{Ue}{\delta E}\cdot \frac{\Delta t}{\tau _{trans}} 
\]
Here $\delta E$ is the mean separation of electron energy levels, 
$Ue/\delta E$ is the number of levels effectivelly contributing to electric 
current, and $\tau _{trans}$ is the mean transmission time required for 
one electron jump from a given level at one side to wherever at opposite 
side. Though quantum jumps can realize in a moment, that moment is random 
and may lie approximately equally anywhere in the interval 
$0<\Delta t$ $<\tau _{trans}$ , 
with $\tau _{trans}$ being the time necessary to accumulate the quantum
transmission probability up to a value $\sim 1$ . Clearly, 
$e/\tau _{trans}$ serves as the mean current per level. Here from one 
gets the tunnel conductance as
\begin{equation}
G=\Delta Q/U\Delta t=e^2\nu \gamma \,\,,\,\,\nu =1/\delta E,\;\gamma
\,=1/\tau _{trans}
\end{equation}
with $\nu $ being electron density of states and $\gamma $ mean jump
probability per unit time.

Of course, any real junction possesses a finite capacity $C$ and thus finite
characteristic time 
\[
\tau _{rel}\approx RC\equiv C/G 
\]
Its physical role can be identified (as usually in RC-circuits) as the
relaxation time of junction charging and the correlation time of thermal
voltage fluctuations (if only suppose that Coulombian interaction between
sides manifests itself in stochastic form). Compare the above defined times: 
\begin{equation}
\tau _{trans}/\tau _{rel}\approx e^2\nu /C=E_C/\delta E
\end{equation}
with $E_C=e^2/C$ being the Coulomb energy. Clearly, (2) is merely number of
levels promoting charge relaxation.

Now let us demonstrate that this ratio may essentially exceed unit, 
\begin{equation}
\tau _{trans}/\tau _{rel}>>1
\end{equation}
even if Coulombian effects are weak in the trivial sense $E_C<<T$ . For
certainty, consider a flat junction with area $S$ , side thicknesses $w$ ,
and with $d$ and $\epsilon $ being the thickness and dielectricity of
isolating barrier, respectively, and use formula 
$C\approx \epsilon S/4\pi d$ . The estimate for the metallic density 
of states is $\nu \approx \mu /N_e$ $\approx Sw/(\hbar v_Fa^2)$ , 
where $\mu $ is Fermi energy, $N_e$ number of metallic electrons, 
$v_F$ Fermi velocity and $a$ is atomic size ( $Sw/N_e=a^3$ 
is volume per one electron), thus 
\[
e^2\nu /C\approx \frac{e^2}{\hbar c}\cdot \frac c{v_F}\cdot \frac{4\pi dw}{%
\epsilon a^2}\approx \frac{dw}{a^2} 
\]
where $c$ is the speed of light (introduced to see that $c/v_F$
overcompensates the fine structure constant) and typical value 
$\epsilon \sim 20$ is substituted.

Obviously, the inequality (3) is satisfied if either $d$ or $w$ (moreover 
if both) a few times exceeds $a$ , i.e. practically always (in this sense
Coulombian effects never could be neglected). Therefore the quantum
probability of some particular electron jump grows inevitably under
influence of relatively fast varying inter-side voltage $u(t)$ , 
$u(t)\sim \sqrt{T/C}$ , produced by many other jumps at the same time 
factually happening in both directions between other levels. Not only a 
moment of jump realization is random but the jump probability itself 
turns be random, that is different kinetic events become entangled.
This may be named quantum (Coulombian) interaction of electron transfers. 
It can not be completely described by one-electron language, but the 
insert of fluctuating voltage gives the quasi one-particle approximation 
which, to some extent, may substitute for a real picture.

\section{Time ratios connected with energy discreteness}

The standard kinetic scheme deals with the tunnel Hamiltonian 
\[
H=H_0+H_{tun}\,\,,\,\,H_{tun}=\sum_{kq}g_{kq}(b_q^{+}a_k+a_k^{+}b_q) 
\]
and with three ansatzes attracted to avoid formal difficulties brought in by
the discreteness of electron energy levels, namely: i) energy spectrum in
sides is so dense that the continuous limit is possible, 
$\sum_k...$ $\rightarrow $ $\int ...\nu (E)dE$ ; 
ii) the Fermi's golden rule 
$p_{kq}(\Delta t)$ $\rightarrow $ $2\pi \Delta t(g_{kq}^2/\hbar )$ 
 $\delta (E_{kq})$ , where $p_{kq}(t)$ is jump probability and 
$E_{kq}$ energy difference between states; 
iii) $g_{kq}^2$ is sufficiently smooth function of $E_{kq}$ . 

Here $\Delta t$ is a time interval necessary to adequately
evaluate jump probabilities for kinetic equations.
This scheme requires the restriction $\Delta t<<\tau _{gold}$ , where 
$\tau _{gold}=2\pi \hbar /\delta E$ . 
However, if we wanted to account for effects
of the voltage fluctuations $u(t)$ we would need at least in 
$\Delta t$ comparable with $\tau _{trans}$ , 
i.e. in the additional condition $\tau _{trans}/\tau _{gold}<1$ . 
But in an adequate model just the opposite relation must be expected, 
\begin{equation}
\tau _{trans}/\tau _{gold}>1
\end{equation}
It is easy seen if note that 
$\tau _{trans}/\tau _{gold}$ $=\delta E/\Delta E $ , 
 $\Delta E\equiv 2\pi \hbar /\tau _{trans}$ , 
where $\Delta E$ is the energy uncertainty associated with 
instability of intra-side electron states. Hence, if the desirable 
condition was valid, then close states would be undistinguishable, 
in other words, electron spectra in sides would undergoe 
mutual rebuilding because of too good transparency of tunnel barrier 
(see also [27] on the relation $R/R_0>1 $ ).

To analyse the ratio (4) more carefully, estimate $\tau _{trans}$ . The mean
transported charge $\Delta Q$ is expressed by 
\[
\Delta Q=e(\Delta N_{+}-\Delta
N_{-})=e\sum_{kq}[f(E_k^{-})-f(E_q^{+})]p_{kq}\,\,,\,\Delta N_{\pm
}=\sum_{kq}f(E_k^{\mp })[1-f(E_q^{\pm })]p_{kq} 
\]
where $\Delta N_{\pm }$ is number of electrons tunneling in left (right)
direction, $E^{\pm }$ are energy levels in the sides, 
 $f(E)=1/[1+\exp (E-\mu )/T]$ , and 
\begin{equation}
p_{kq}=p_{kq}(\Delta t,U)\approx 4g_{kq}^2\sin ^2(E_{kq}\Delta t/2\hbar
)/E_{kq}^2\,\,\,,\,E_{kq}=E_q^{+}-E_k^{-}-eU
\end{equation}
is tunneling probability evaluated by ordinary perturbation theory. The
corresponding low field conductance 
\[
G=[\Delta Q/U\Delta t]_{U\rightarrow 0}=e^2\sum_{kq}[-\partial
f(E_k)/\partial E_k]p_{kq}(\Delta t,0)/\Delta t 
\]
turns into Eq.1 with 
\begin{equation}
\gamma =1/\tau _{trans}=\sum_qp_{kq}(\Delta t,0)/\Delta t\approx 2\pi g^2\nu
/\hbar ,\,\,\,(E_k,E_q\approx \mu )
\end{equation}
and $g$ being characteristic magnitude of $g_{kq}$ . Combining this relation
with Eq.1 one obtains 
\begin{equation}
\frac{\tau _{trans}}{\tau _{gold}}=\frac{e^2}{2\pi \hbar G}=\frac R{R_0}%
\simeq \left( \frac{\delta E}{2\pi g}\right) ^2\,\,\,,\,\,R_0=\frac{2\pi
\hbar }{e^2}\sim 20\,\text{kOhm}
\end{equation}

Hence, the condition for really weak interaction is the smallness of
transfer matrix elements as compared with the energy level spacing.

\section{Discreteness, phase decoherence and fluctuations of quantum
transfer probabilities}

Since the wish $\tau _{trans}/\tau _{gold}<1$ was invoked by the continuous
spectrum approximation, we may suspect that the discreteness must be
involved in an evident form to describe the fluctuations of transfer
probabilities, while the perturbation theory is still applicable. Consider
the picture when quantum transfers from a fixed level ''k'' at one side to
any level ''q'' at another side are influenced by a fast fluctuating field
(FFF), here the voltage noise $u(t)\,$ , in its turn induced by electron
jumps beyond our attention. Thus we use quasi one-electron picture, basing
on the ansatz that in many-electron surroundings with a sufficiently rich
energy spectrum $u(t)$ behaves as a random process. The modern theory of
quantum chaos gives powerful support for this statement [33,34] (although
taking in mind that in specific systems coherent collective charge
oscillations are possible instead of stochasticity and relaxation).

In this section we may consider thermodynamical equilibrium taking $U=0$ .
First introduce the randomly accumulating diffusive phase shift 
\[
\,\varphi (t)=\frac e\hbar \int_0^tu(t^{\prime })dt^{\prime }
\]
Instead of (5), the standard Shroedinger equations for transfer amplitudes
yield 
\begin{equation}
p_{kq}\approx \left| A_{kq}\right| ^2\,,\,\,\,A_{kq}\equiv \frac{g_{kq}}\hbar
\int_0^{\Delta t}\exp (iE_{kq}t/\hbar )\,Z(t)dt\,\,,\,\,\,Z(t)=\exp
[i\varphi (t)]
\end{equation}
Clearly, now $p_{kq}$ become random values governed (parametrically excited
or damped) by the phase shift in its turn produced by fluctuating potential
difference between ''in'' and ''out'' states. Further, introduce the phase
correlation function, the phase correlation time and corresponding
energetical measure of quantum coherence by  
\begin{equation}
K(t_1-t_2)=\left\langle Z(t_1)Z^{*}(t_2)\right\rangle \,\,\,,\,\tau
_{phase}=\int_0^\infty \left| K(\tau )\right| d\tau \,\,,\,\,\varepsilon
_{coh}=2\pi \hbar /\tau _{phase}
\end{equation}
where angle brackets denote averaging with respect to FFF.

As demonstrated below, expectedly $\tau _{phase}$ is rather short as
compared with $\tau _{trans}$ and naturally limited by $\tau _{rel}$ .
Therefore under the integral in (8) $Z(t)$ can be treated as complex shot
noise. Consequently, at $\Delta t>>\tau _{rel}$ the transfer amplitudes 
 $A_{kq}$ behave approximately as complex Brownian paths, and regardless 
of details of $Z(t)$'s statistics we have reasons to write 
\begin{equation}
\left\langle p_{kq}^2\right\rangle =\left\langle \left| A_{kq}\right|
^4\right\rangle \approx 2\left\langle \left| A_{kq}\right| ^2\right\rangle
^2\,=2\left\langle p_{kq}\right\rangle ^2\,\,,\,\,\left\langle
p_{kq},p_{kq}\right\rangle \approx \left\langle p_{kq}\right\rangle ^2
\end{equation}
Here the Malakhov's cumulant brackets are used,
\[
\left\langle x,y\right\rangle \equiv \left\langle xy\right\rangle
-\left\langle x\right\rangle \left\langle y\right\rangle 
\]

This is our first finding: if considered at time intervals of order of the
actual transition time the quantum amplitudes and probabilities may become
100\% uncertain due to the phase decoherence (phase diffusion) forced by
FFF. The second is that the mean probabilities grow linearly with time: 
\[
\left\langle p_{kq}\right\rangle \approx \Delta t(g/\hbar )^2\int K(\tau
)\exp (iE_{kq}\tau /\hbar )d\tau \propto \Delta
t\,\,,\,(\,E_{kq}=E_q^{+}-E_k^{-}) 
\]
i.e. FFF remove a need in the golden rule to have uniform growth of
probabilities even at $\Delta t>>\tau _{gold}$ . Notice for the next that
according to this formula now $\varepsilon _{coh}$ 
 (instead of $2\pi \hbar /\tau _{gold}$ ) serves as the width of 
energy region available from a given level (''k'' or ''q'').

But most interesting subject is the summary transition rate 
$\gamma =p/\Delta t$ , with  $p=p_k$ , 
\begin{equation}
p_k\equiv \sum_qp_{kq}=\int \int_0^{\Delta t}\Gamma
(t_1-t_2)Z(t_1)Z^{*}(t_2)dt_1dt_2\,\,,\,\,\Gamma (\tau )=\sum_q(\frac{g_{kq}}%
\hbar )^2\exp (i\tau E_{kq}/\hbar )
\end{equation}
Here the kernel $\Gamma (\tau )$ represents the discreteness. Its analytical
properties are of much importance for all the theory . In the continuous
approximation provided ansatzes i)-iii) one would have 
 $\Gamma (\tau )=$ $\delta (\tau )/\tau _{trans}$ , 
and $p$ would be definitely nonrandom, $p=\Delta t/\tau _{trans}$ . 
In reality, because of the discreteness this
kernel is quite non-local and does not decay completely at arbitrary long
time. For visuality only, let us choose equidistant spectrum at the
right-hand side, $E_q^{+}-E_k^{-}=$ $n\delta E+\varepsilon $ ( $n$ is
integer, $\varepsilon \sim \delta E$ ), then  
\begin{equation}
\Gamma (\tau )=\frac 1{\tau _{trans}}\exp (i\varepsilon \tau /\hbar
)\sum_n\delta (\tau -n\tau _{gold})
\end{equation}
Now the third important point is easy seen : 
if $\varepsilon _{coh}>\delta E$  then the mean probability 
 $\left\langle p\right\rangle \approx $ $\Delta t/\tau _{trans}$ 
 practically coinsides with what is given by usual kinetics,
even in spite of formal violation of the golden rule (in this sense, FFF
effectively expand applicability of usual scheme).

And the main fourth point is that the phase decoherence produced by FFF if
combined with the discreteness results in randomness of the summary quantum
jump probabilities. Indeed, due to the possibility to consider $A_{kq}$'s as
Brownian walks we obtain 
\begin{equation}
\left\langle p,p\right\rangle \approx \int ...\int_0^{\Delta t}\Gamma
(t_1-t_2)\Gamma (t_3-t_4)K(t_1-t_4)K(t_3-t_2)dt_1...dt_4
\end{equation}
Here under the same condition $\varepsilon _{coh}>\delta E$ only the regions 
$t_1\approx t_4,$ $t_3\approx t_2$ are significant but many delta functions
from (12) contribute. The resulting transfer probability variance is 
\begin{equation}
\left\langle p,p\right\rangle \approx \frac{\Delta t^2}{\tau _{trans}^2\tau
_{gold}}\int \left| K(\tau )\right| ^2d\tau \approx \frac{\tau _{phase}}{%
\tau _{gold}}\left\langle p\right\rangle ^2
\end{equation}
(we took into account that ''width'' of delta functions determined by the
inverse width of whole energy band is wittingly smaller 
than $\tau _{phase}$ ).

If inequality (4) was inverted because of too small $\delta E$ the
expression (13) would formally turn into zero, but in fact it always remains
nonzero under any realistic slightly non-equidistant spectrum. Nevertheless,
we may predict that violation of (4), as well as of (3), leads to supression
of the probability fluctuations. Oppositely, large 
 $\delta E>\varepsilon _{coh}$ means most strong fluctuations with variance 
 $\sim \left\langle p\right\rangle ^2$ or greater. 
But thats are accompanied by decreasing
correlations between transfers from different levels:  
\begin{equation}
\left\langle p_{k_1\,,\,}p_{k_2}\right\rangle \approx \left\langle
p_{k_1}\right\rangle \left\langle p_{k_2}\right\rangle \frac 1{\tau _{gold}}%
\int \exp [i(E_{k_1}^{-}-E_{k_2}^{-})\tau /\hbar ]\left| K(\tau )\right|
^2d\tau 
\end{equation}
Evidently, while at $\varepsilon _{coh}>\delta E$ the currents from many
levels fluctuate concordanly, at $\varepsilon _{coh}<\delta E$ even close
levels inject independently one on another. Besides, even 
 $\left\langle p\right\rangle $ 
 become sensible to the energy shift $\varepsilon $ in (12),
therefore, this extreme case should be carefully analysed with accounting
for realistic (uncommensurable) discrete energy spectra (and better out of
the frame of the quasi one-particle picture).

To end this section, estimate the phase correlation time $\tau _{phase}$ 
(what may be named also phase decoherence time). Notice that $K(\tau )$ is
nothing but characteristic function of the phase. In principle, it is
determined just by the transfer statistics, in a complicated self-consistent
picture. Since the latter now is out of our look, we confine ourselves by
rough reasonings. For instance, at $E_C$ $<<$ $T$ it is likely natural to
treat $u(t)$ as the Ornstein-Uhlenbeck random process, then 
\[
K(\tau )\approx \exp \left[ \frac TC\left( \frac e\hbar \tau _{rel}\right)
^2\{1-\frac \tau {\tau _{rel}}-\exp (-\frac \tau {\tau _{rel}})\}\right] 
\]
The corresponding decoherence time is 
 $\tau _{phase}\sim $ $(\hbar /e)\sqrt{C/T}$ $(<<\tau _{gold})$ . 
Perhaply, this is low bound for it (too rigid in
the sense that it does not include the conductance $G$ ). At $E_C\sim T$ 
(what qualifies small junctions) the charge quantization is essential and
the better model for $\varphi (t)$ is infinitely divisible random walk (on
this subject see [37]) formed with rare increments by 
 $\Delta \varphi =$ $(e/\hbar )(\pm e/C)\theta $ , 
where $\theta $ is random duration of charged
stay distributed with some probability density $W(\theta )$ and total time
fraction $\sim 1$ . The suitable expression is 
 $W(\tau )=$ $\tau _{rel}^{-1}\exp (-\tau /\tau _{rel})$ 
what corresponds to the characteristic function 
\[
\Xi (\eta ,\tau )\equiv \left\langle \exp [i\eta \int_0^\tau
u(t)dt]\right\rangle \approx \exp \left\{ \frac{|\tau |}{\tau _{rel}}%
\int_0^\infty [\cos \left( \frac{\eta e}C\theta \right) -1]W(\theta )d\theta
\right\} 
\]
(here the Levy-Khinchin representation [37] was applied). 
Taking $\eta =e/\hbar $ and assuming $2\pi R/R_0$ $>1$ , obtain 
 $K(\tau )=$ $\Xi (e/\hbar ,\tau )\approx $ $\exp (-|\tau |/\tau _{rel})$ . 
Thus, in this oppositely
extreme case $\tau _{phase}$ may be of order of $\tau _{rel}$ (likely
representing upper bound for $\tau _{phase}$ ) and possibly larger 
than $\tau _{gold}$ .

\section{Low-frequency conductance fluctuations}

In this section we return to externally driven junction, so the total
voltage will be $U(t)=U+u(t)$ with the applied voltage $U$ interpreted as
its average value.

Of course, fluctuations of electron jump probabilities eventually result in
more or less analogous conductance fluctuations. The conductance is implied
as $\Delta Q/U\Delta t$ where $\Delta Q$ is the conditional quantum average
value of transported charge taken under a fixed realization of $u(t)$ . The
charge transport depends also on occupancies of energy levels. If $U=0$ then 
$\Delta Q$ , by its definition, must turn into zero regardless of $u(t)$ .
This means the existence of some natural statistical correlations between
voltage noise and occupancies whose accurate description would need in a
self-consistent many-electron picture. To avoid this difficulty, let us work
up the storage 
\begin{equation}
\Delta Q=e\sum_{kq}[f(E_k^{-})-f(E_q^{+})]\left| A_{kq}\right| ^2
\end{equation}
making time integration by parts in the amplitudes as if $u(t)$ was absent,
to transform (16) into 
\begin{equation}
\Delta Q=e^2U\int \int_0^{\Delta t}\Lambda (t_1-t_2)\exp
[-ieU(t_1-t_2)/\hbar ]Z(t_1)Z^{*}(t_2)dt_1dt_2
\end{equation}
where new kernel is introduced,  
\begin{equation}
\Lambda (\tau )=\sum_{kq}(\frac{g_{kq}}\hbar )^2\frac{f(E_k^{-})-f(E_q^{+})}{%
E_q^{+}-E_k^{-}}\exp [i\tau (E_q^{+}-E_k^{-})/\hbar ]
\end{equation}
This form is consistent with the detailed balance $\Delta Q(U=0)=0$ and
hence may serve for estimates.

In general, if electron spectra in sides are much wider than $T$ then 
at $U<T/e$ we believe that, approximately, 
i) conductivity obeys Ohmic law and
ii) its relative fluctuations do not depend on $U$ . Then Eqs. 17-18 can be
simplified by means of linearization into 
\begin{equation}
\Delta Q\approx e^2U\sum_k[-\partial f(E_k^{-})/\partial E_k^{-}]p_k
\end{equation}
with $p_k$ defined by (11) . Naturally, the conductance fluctuations
essentially depend on relation between decoherence and level spacing. At
sufficiently small phase decoherence time (''large'' junction, widely
correlated jumps from different levels), the Eq.15 helps easy obtain 
\begin{equation}
\left\langle G,G\right\rangle \sim \frac{\delta E}T\left\langle
G\right\rangle ^2\,\,\,,\,\,(\tau _{phase}<<\tau _{gold})
\end{equation}
At large decoherence time (''small'' junction, non-correlated jumps)
conductance fluctuations are very sensible to concrete structure of energy
spectra in sides, first of all to the degree of their relative
commensurability. Omitting the details, the result is that the
relative conductance variance may prove to be anywhere in the interval  
\begin{equation}
\frac{\tau _{phase}}{\tau _{gold}}\cdot \frac{\delta E}T<\frac{\left\langle
G,G\right\rangle }{\left\langle G\right\rangle ^2}<1\,\,,\,\,(\tau
_{phase}>\tau _{gold})
\end{equation}
i.e can achieve as great magnitude as $\sim 1$ . In this case, accurate 
accounting for fluctuations of the occupancies may be especially 
important. 

\section{Comparison with experiment}

The good experimental illustration for permissibly exciting properties of
quantum transfers was brought in [36] where 1/f-noise in the cermet
(granular composite) $Ni$-nanoparticles $(25\%)$-$Al_2O_3$was investigated.
In this system the parameters of a typical elementary tunnel junction formed
by neighbouring metal particles are $\delta E\approx 0.2\,$meV, 
 $d\approx 2\,$nm, $C\approx 5\cdot 10^{-6}\,$cm, 
 $E_C\sim T$ (at room temperature), and $R\approx 30\,$MOhm, which mean 
that $\tau _{gold}\approx 3\cdot 10^{-11}\,$s , 
 $\tau _{rel}\approx 1.5\cdot 10^{-10}\,$s , 
 $\tau _{trans}/\tau _{rel}=E_C/\delta E\approx 200\,$ and 
 $\tau _{trans}\approx 3\cdot 10^{-8}\,$s . 
Both the inequalities (3) and (4) are well satisfied, thus giving us all
grounds (see also Discussion) to suspect that 1/f-noise could be attributed
to quantum Coulombian interactions, in the above mentioned sense (possibly
with a contribution by electron-phonon processes). 
Moreover,  $\tau _{gold}\,$ is even noticably smaller than 
 $\tau _{rel}$, thus indicating, from our point of view, the possibility 
of highly maintained conductance fluctuations.

In fact, such cermet is characterized by giant 1/f conductance noise with
relative spectrum density $S_{\delta G}(f)\approx \alpha /N_gf\,$ where 
 $\alpha \approx 6\cdot 10^{-3}\,$ and $N_g\,$ is the number of metal 
particles in a sample. In view of $E_C\sim T\,$, $N_g\,$ approximately 
represents the number of active (simultaneously transported) electrons 
[36], hence that is almost standard noise with classical Hooge 
constant ($\alpha =2\cdot 10^{-3}$) [1,2,5]. 
It corresponds to the $S_{\delta G}(f)\sim \alpha /f\,$ noise
in separate elementary junction, demonstrating 
at least $\sim 100\%\cdot \sqrt{\alpha \ln (1\,\text{s}/\tau _{rel})}$ , 
i.e. $\sim 40\%$, uncertainty of its conductance.

But most beautiful observation [36] was the visible 1/f-noise sensibility to
the discreteness of electron energy spectra in metal granules. When applied
voltage per elementary junction exceeds $\delta E/e\,$ and thus
corresponding current exceeds $e/\tau _{trans}$ (see Eq.1), 
then quadratic (low-field) dependence of variance of voltage 1/f noise 
on bias current transforms into ''non-Ohmic'' linear one (although mean 
conductance obeys Ohm law up to $\sim T/\delta E\sim 100\,$ times 
larger current).

The experimental appearance of the factor $\delta E/e$ is evident 
manifestation of the role of discreteness and it itself gives 
clear confirmation of our aproach to 1/f-noise in this system. 
At the same time, this means that the hypothesis ii) (in previous 
section) fails, i.e. linearized expression (19) becomes invalid if 
applied to fluctuations, and the estimates (20)-(21) should be multiplied 
by a decreasing factor $D(U)$ ($D(0)=1$ , $D(U)<1$ ). 
Indeed, as we underlined, any transfer from one side embraces 
$\approx \varepsilon _{coh}/\delta E\,$ levels at another
side. If this number is smaller than the number of charge transporting
levels, $eU/\delta E$ , then the latters act as uncorrelated quantum
channels. Hence, $\Delta Q \,$ variance may become linear function of 
$U$ being proportional to the number of channels, what corresponds in 
small junctions to $D\sim \delta E/eU\,$ at $eU> \delta E $ . 

Definite formal support for this euristic reasoning (literally appropriate 
at zero temperature only) could be found already in the frame of quasi 
one-electron picture, if consider full non-linear dependence (17) of 
$\Delta Q $ on $U$ . The decreasing factor what follows 
from the Eqs.17-18 is 
\begin{equation}
D(U)=S(eU)/S(0)\,\,,\,\,S(E)\equiv \int K_\Lambda (\tau )\exp (-iE\tau
/\hbar )\left[ \int K(\theta )K(\tau -\theta )d\theta \right] d\tau 
\end{equation}
where the correlation function $K_\Lambda (\tau )$ is defined by 
\begin{equation}
K_\Lambda (\tau )=\frac 1{2\Delta t}\int_{-\Delta t}^{\Delta t}\Lambda (t+%
\frac \tau 2)\Lambda ^{*}(t-\frac \tau 2)dt
\end{equation}
In case of small junctions the averaging over an ensemble of probable
electron spectra must be added what automatically takes place in cermets. In
theory, we need in some statistics of energy levels (see [34]
on the known variants). At present, notice only that the absense of a rigid 
measure of local spacing of energy levels (except the mean 
value $\delta E$ as related to whole spectra only) might results 
naturally in a weak dependence of 
$K_\Lambda (\tau )$ on $\tau $ characterized by logarithm 
$\ln (\tau _{gold}/\tau )$ which implies just the inverse proportionality 
$D(U)\sim \varepsilon _{coh}/eU$ at $eU>\varepsilon _{coh}$ . 

\section{Discussion and resume}

The principal peculiarity of above considered fluctuations of quantum
transfer probabilities and corresponding conductance fluctuations is that
their relative measure seems independent on the time under observation, 
 $\Delta t$ . It looks as if conductance undergoe 
fluctuations with non-decaying statistical correlations. 
Outwardly thats resemble static
fluctuations investigated in the theory of disordered conductors (so-called
universal conductance fluctuations) [34]. But by essence these are different
things: one starts from the decoherence just when other finishes at it. 

Though our analysis was limited by times $\Delta t\sim \tau _{trans}$ ,
there is a feeling that similar statistics expands to arbitrarily longer
time scales. Formal proof that it is really true, attributing to Hamiltonian
models of quantum channels (with ''weak links'' like tunnel barriers)
subjected to FFF (Coulombian or magnetical), will be done separately. For
the present, semi-formal arguments are: i) all what was obtained results
from the trivial rule that in general one should manipulate with quantum
amplitudes to find quantum probabilities as a final product; ii) all what
was obtained results from ''fast'' noise and phase decoherence, but not from
some causal correlations which could not be continued to longer time. We
believe that a self-consistent analysis connecting FFF and electron
transport statistics will result in some slow (nearly logarithmic)
dependence of $\left\langle G,G \right\rangle $ on the observation time
reflecting non-static character of the fluctuations and described by 1/f
spectrum, instead of $\delta (f)$ having the same formal dimensionality 
(such dependence may begin already as a weak $\Delta t $ dependence of 
the correlator (23)).

An accurate approach to a tunnel junction would deal with non-equilibrium
steady state governed by the Hamiltonian 
\[
H=H_0+H_{tun}\,,\,\,H_0=H_{-}+H_{+}+H_C-U\Delta Q\, 
\]
where $H_C$ describes inter-side Coulombian interaction, $\Delta Q$ is the
operator of transported charge, and $H_{\pm }$ describe two sides with their
leads serving also as thermal baths ensuring relaxation to equilibrium
occupancies. Like in practice, in theory it is rather hard to ''solder
leads'', especially if wishing to avoid appeals to standard kinetic schemes.
But, regardless of tehnical difficulties, we may state that none processes
in leads could influence the formation of non-diagonal (inter-side) elements
of the operator $\Delta Q $ since thats are determined by the tunneling
itself and charging of tunnel barrier only. Therefore, the better relaxation
in leads the more grounds we have to extend the Eq.16 to arbitrary time
intervals (now with $\left| A_{kq} \right| ^2 $ representing the number of 
passing electrons). The comparatively non-principal corrections to be 
performed are accounting for fluctuations of occupancies and including 
conductance fluctuations into higher-order statistics 
of FFF (voltage noise). 

To resume, we demonstrated that if do not neglect the actual quantum 
discreteness when constructing kinetic models of transport processes then 
possible strong sensibility of quantum transfer amplitudes and 
probabilities to fast fluctuating fields becomes visible (in particular, 
created by the transfers themselves) which may result in 
fundamental 1/f type low-frequency fluctuations of transport rates. 
Hence, the now reigning quantum kinetic models ask for evident general
comments. All come from the well known Pauli's kinetic master equations. In
[35] Van-Hove developed its formal groundation under so called 
$\lambda ^2t$ -limit. But, with no doubts, this theory does not foresee 
anything like 1/f noise. What is the matter, the question arises naturally 
(some comments on this issue were suggested in [24]). We hope that our above 
consideration highlights the possible principal answer: in fact the 
Van-Hove's formalism supposes the limit $\delta E\rightarrow 0$ (i.e. 
the continuous spectrum idealization) be performed before the limit 
$g\rightarrow 0$ with $g$ ($g=\lambda $) representing (as above) the 
magnitude of weak interactions. Thus the influence of quantum discreteness 
onto statistics of quantum jumps (between eigenstates of unperturbed 
Hamiltonian) is lost. We think that it will be not hopeless to 
properly improve the present kinetics (see Introduction).

\,\,

ACKNOWLEDGEMENTS

\,\,

I acknowledge Dr. Yu.V.Medvedev and participants of his seminar at the
Department of kinetic properties of disordered and nonlinear systems in
DonPTI NAS of Ukraine for support, stimulating criticism and helpfull
discussions.

\,\,

REFERENCES

\,\,

1. P.Dutta, P.Horn, Rev.Mod.Phys., 53(3), 497 (1981).

2. F.N.Hooge, T.G.M.Kleinpenning, L.K.J.Vandamme, Rep.Prog.Phys., 44, 481
(1981).

3. M.B.Weissman, Rev.Mod.Phys., 60, 537 (1988).

4. M.B.Weissman, Rev.Mod.Phys., 65, 829 (1993).

5. G.N.Bochkov and Yu.E.Kuzovlev. UFN, 141, 151 (1983), English translation:
Sov.Phys.-Usp., 26, 829 (1983).

6. B.Raquet, J.M.D.Coey, S.Wirth and S. Von Molnar, Phys.Rev., B59, 12435
(1999).

7. J.G.Massey and M.Lee, Phys.Rev.Lett., 79, 3986 (1997).

8. M.J.C. van den Homberg, A.H.Verbruggen, P.F.A.Alkemade, S.Radelaar,
E.Ochs, K.Armbruster-Dagge, A.Seeger and H.Stoll, Phys.Rev., B 57, 53 (1998).

9. A.Ghosh, A.K.Raychaudhuri, R.Streekala, M.Rajeswari and T.Venkatesan,
Phys.Rev., B 58, R14666 (1998).

10 X.Y.Chen, P.M.Koenrad, F.N.Hooge, J.H.Wolter and V.Aninkevicius,
Phys.Rev., B 55, 5290 (1997).

11. G.M.Khera and J.Kakalios, Phys.Rev., B 56, 1918 (1997).

12. M.Gunes, R.E.Johanson and S.O.Kasap, Phys.Rev., B 60, 1477 (1999).

13. K.M.Abkemeier, Phys.Rev., B 55, 7005 (1997).

14. G.Snyder, M.B.Weisman and H.T.Hardner, Phys.Rev., B 56, 9205 (1997).

15. A.Lisauskas, S.I.Khartsev and A.M.Grishin, Studies of 1/f-noise in $%
La_{1-x}M_xMnO_3$ (M=Sr,Pb) epitaxial thin films, in J.Low.Temp.Phys. as
MOS-99 Proceedings.

16. A.Lisauskas, S.I.Khartsev, A.M.Grishin and V.Palenskis, Electrical noise
in ultra thin giant magnetoresistors, in Mat.Res.Soc.Proc. Spring-99 Meeting.

17 B.I.Shklovskii and A.L.Efros. Electronic properties of doped
semiconductors, Springer-Verlag. Berlin, 1984.

18. V.I.Kozub, Solid State Commun., 97, 843 (1996).

19. V.Podzorov, M.Uehara, M.E.Gershenson and S.-W.Cheong, lanl arXiv
cond-mat/9912064.

20. M.Viret, L.Ranno and J.M.D.Coey, Phys.Rev., B 55, 8067 (1997).

21. Yu.E.Kuzovlev and G.N.Bochkov. On the nature and statistics of
1/f-noise. Preprint No.157, NIRFI, Gorkii, USSR, 1982.

22. Yu.E.Kuzovlev and G.N.Bochkov.Izv.VUZov.-Radiofizika, 26, 310 (1983),
transl. in Radiophysics and Quantum Electronics (RPQEAC, USA), No 3 (1983).

23. G.N.Bochkov and Yu.E.Kuzovlev. Izv.VUZov.-Radiofizika, 27, 1151 (1984),
transl. in Radiophysics and Quantum Electronics (RPQEAC, USA), No 9 (1984).

24. G.N.Bochkov and Yu.E.Kuzovlev. On the theory of 1/f-noise. Preprint N
195, NIRFI, Gorkii, USSR, 1985.

25. Yu.E.Kuzovlev, lanl arXiv cond-mat/9903350 .

26. N.S.Krylov. Works on the foundations of statistical mechanics. Princeton
U.P., Princeton, 1979 (Russian original book published in 1950).

27. Yu.A.Genenko and Yu.M.Ivanchenko, Teor.Mat.Fiz., 69, 142 (1986). 

28. Yu.E.Kuzovlev, Zh. Eksp. Teor. Fiz, 94, No.12, 140 (1988), transl. in
Sov.Phys.-JETP, 67 (12), 2469 (1988).

29. Yu.E.Kuzovlev, Phys.Lett., A 194, 285 (1994).

30. Yu.E.Kuzovlev, Zh. Eksp. Teor. Fiz, 111, No.6, 2086 (1997), transl. in
JETP, 84(6), 1138 (1997).

31. P.Bak. Self-organized criticality: why nature is complex. Springer, N-Y,
1996.

32. J.M.D.Coey, M.Viret and S. von Molnar, Adv.Phys., 48, 167 (1999).

33. G.Casati and B.Chirikov. Fluctuations in quantum chaos. Preprint, Budker
Inst. of Nuclear Physics SB RAS, 1993.

34. C.W.J.Beenakker, Rev.Mod.Phys., 69, N 3, 731 (1997).

35. L.Van Hove, Physica, 21, 517 (1955).

36. J.V.Mantese, W.I.Goldburg, D.H.Darling, H.G.Craighead, U.J.Gibson,
R.A.Buhrman and W.W.Webb, Solid State Commun., 37, 353 (1981).

37. W.Feller, Introduction to probability theory and its applications, John
Wiley, N-Y, 1966.

38. Yu.A.Genenko and Yu.M.Ivanchenko, Phys.Lett., 126, 201 (1987). 

\end{document}